\documentclass[prl,aps,nofootinbib,showkeys,showpacs,twocolumn,floatfix]{revtex4}   
\usepackage{epsfig}  
\usepackage{graphicx}  
\usepackage{color}  
\begin{document}  
  
\title{Thermodynamics of small superconductors with fixed particle number}

\author{Danilo Gambacurta}  
 \email{gambacurta@ganil.fr}  
\affiliation{GANIL, CEA/DSM and CNRS/IN2P3, Bo\^ite Postale 55027, 14076 Caen Cedex, France}  
\author{Denis Lacroix} \email{lacroix@ganil.fr}  
\affiliation{GANIL, CEA/DSM and CNRS/IN2P3, Bo\^ite Postale 55027, 14076 Caen Cedex, France}  

\begin{abstract}  
The Variation After Projection approach is applied for the first time to the pairing hamiltonian to describe
the thermodynamics of small systems with fixed particle number.  The minimization of the free energy is made 
by a direct diagonalization of the entropy. The Variation After Projection applied at finite temperature provides a perfect reproduction of the exact canonical properties of odd or even systems  from very low to high temperature.

\end{abstract}  

\maketitle  
  
\section{Introduction}  
Recent progress in single-electron tunneling spectroscopy have revealed the
persistence of pairing effect even at very small number of particles
\cite{Bra98}. The tremendous experimental work in ultra-small metallic grains
\cite{Von01} has enabled to systematically investigate the transition from large
systems, the bulk limit, up to very small systems. By varying the number of
particles, thermal excitations or adding external magnetic fields, the smearing
of superfluid-to-normal phase transition, the survival of pairing correlations,
the odd-even staggering \cite{Bra99,Bal98} and/or possible re-entrant
effects \cite{Dil00} have been carefully analyzed.   These studies  have
underlined the importance of finite size effect on pairing correlations and the
necessity to develop theories beyond the Bardeen-Cooper-Schieffer (BCS) or the
Hartree-Fock-Bogoliubov (HFB) ones that properly account for particle number
conservation. Some of these studies are at the crossroad with nuclear physics
where systems contain very few to several hundreds of nucleons \cite{Rin80} and
some of the approaches that are used nowadays to deal with particle number
conservation, like projection techniques \cite{Ben03,Fer06} have been imported
in condensed matter \cite{Bra98}. In this case, improvement beyond the BCS
and/or HFB, is obtained by considering a state with good particle $| \Psi_N
\rangle = P_{N} | \Phi_0 \rangle$, where $P_N$ is the projector on $N$
particles  while $| \Phi_0 \rangle$ denotes a quasi-particle (BCS or HFB)
state. The explicitly breaking of the symmetry, the $U(1)$ one in the present
case, allows to grasp the physics of pairing while its restoration is required
to describe the onset of pairing in very small systems (see for instance Fig. 1
of ref. \cite{Hup11}).

A natural extension of this approach able to provide a canonical description of
finite system at thermal equilibrium has been proposed already some times ago
\cite{Ese93} by considering a many-body projected density $\hat D_N$ written as
(see also \cite{Bal99}): \begin{eqnarray} \hat D_N = \frac{1}{Z} {\hat P}_N
\exp(-\beta \hat h)  {\hat P}_N , \label{eq:dn} \end{eqnarray} where $Z={\rm Tr}
({\hat P}_N \exp(-\beta \hat h)  {\hat P}_N)$, $\beta = 1/(k_B T)$, and $\hat h$
is the quasi-particle effective BCS or HFB hamiltonian. In view of the complexity
of this approach, approximations or alternative theories have been proposed.
In ref. \cite{Ross1994}, a general projection formalism was developed and largely applied in the
static-path-approximation. The problem of particle 
number projection at finite temperature was also addressed  in the context of the
thermofield dynamic \cite{Tanabe} but no applications have been done till now.
Starting from a mean-field plus pairing description
in the Grand-Canonical ensemble, several improvements of increasing complexity
have been proposed to correct from particle number explicit non-conservation.
Along this line, a Modified BCS theory \cite{Din03} has been introduced where
part of the statistical fluctuations is directly incorporated in the
quasi-particle transformation. This approach has been further improved by
extending the Lipkin-Nogami approach to finite temperature, projecting onto good
particle number after variation or adding quantum fluctuation associated to RPA
modes \cite{Din08}. Note however that its justification and applicability especially at high
temperature remain to be clarified \cite{Pon05}. On the other hand, starting
from a functional integral formulation and treating approximately the collective
fluctuations around the mean-field path, is shown to provide a suitable tool
over a wide range of temperatures but breaks down at very low temperature
\cite{Ros97}. An approximate scheme to deal with quantal fluctuations consists
in the use of a Grand-Canonical plus a parity-projected technique \cite{Bal98,
Dil00,Delft96} which allows to describe qualitatively odd-even effects but still
suffers from of abrupt and/or spurious phase transitions \cite{Von01}. Even in
very schematic models \cite{Ric64}, unless an exact treatment is made either by
direct diagonalization \cite{Sum07}  or by quantum Monte-Carlo techniques
\cite{Van06}, a canonical finite-T method based on mean-field theory and 
valid at arbitrary small or high temperature remains problematic and 
appears as a  challenge in this field \cite{Von01}. 

While the results presented in ref. \cite{Ese93} were very promising, this
method has never been applied due to its complexity. Here, we apply for the
first time the method proposed in ref. \cite{Ese93} to the Richardson
hamiltonian at thermal equilibrium and show that this approach provides a proper
description of both thermal and quantal fluctuations from very low to high
temperature. The canonical description of a quantum finite system can be
obtained  by minimizing the Helmholtz free energy $F$
\begin{equation}
\delta F = \delta ( {\rm Tr} [\hat H \hat D_N] - T S) = 0 , \label{eq:varia}
\end{equation}
where $S$ denotes the entropy associated to the projected density (\ref{eq:dn}), 
i.e. $S = -k_B{\rm Tr}(\hat D_N \ln \hat D_N)$. The approach is applied to
the pairing Hamiltonian written as \cite{Ric64}:
\begin{eqnarray}\label{Hamiltonian}
\hat H = \sum_{i,\sigma = \pm} (\varepsilon_i - \sigma \mu_B B)\hat c^\dagger_{i\sigma}\hat c_{i\sigma}  
-G \sum_{i, j} \hat c^\dagger_{i,+} \hat c^\dagger_{i,-} \hat c_{i,-} \hat c_{i,+},
\end{eqnarray}
where $B$ is an external magnetic field. For not too big systems, thermodynamic
quantities can be studied in different statistical ensembles without
approximation by direct diagonalization of the Hamiltonian in different
seniority spaces\cite{Sum07}. 

\begin{figure}[htbp]  
\includegraphics[width=8.cm]{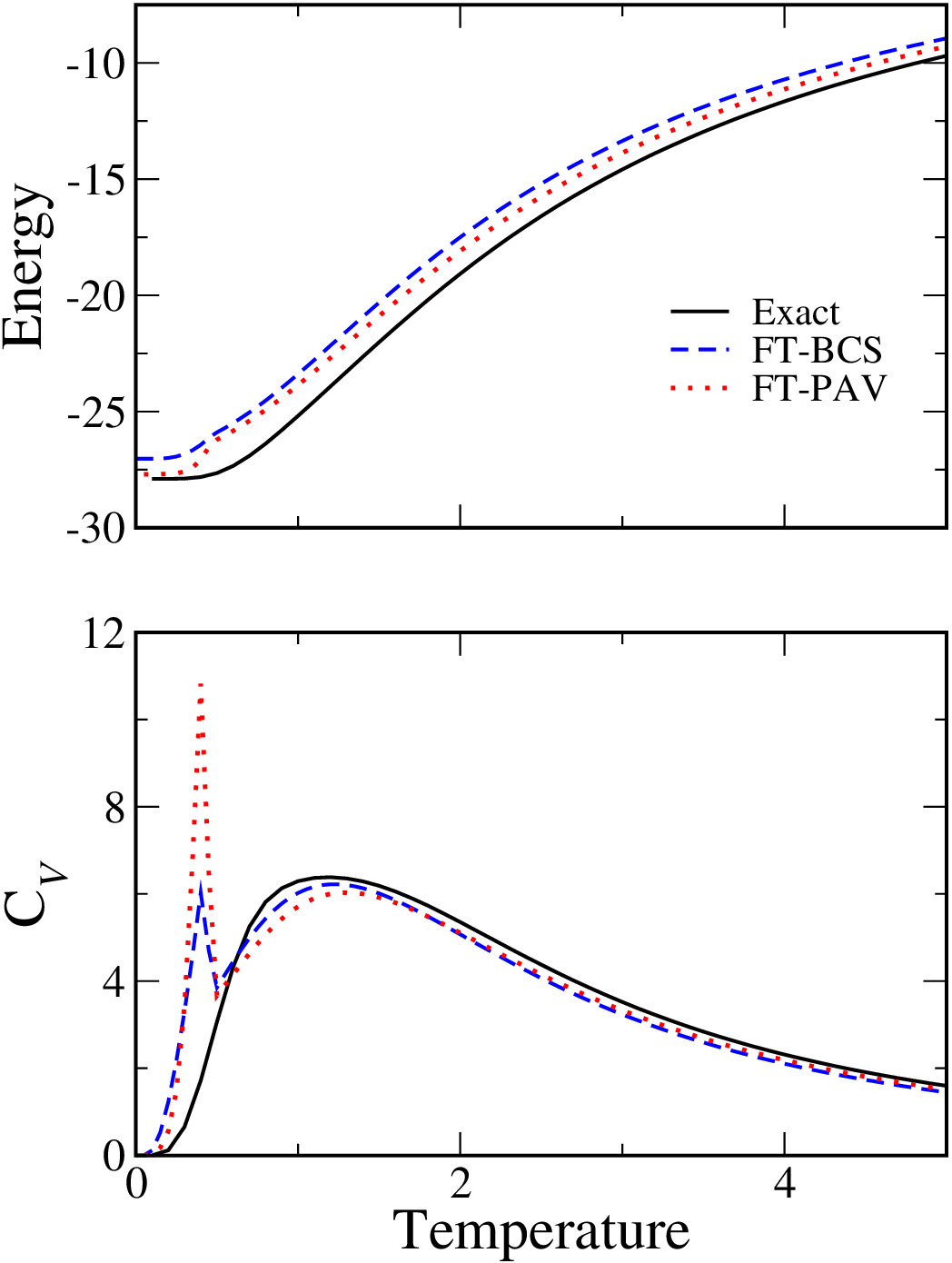}  
\caption{Evolution of the energy (Top) and heat capacity (Bottom) obtained with
the FT-BCS (dashed line), PAV (filled circles) 
and exact solution (thick line) for a system of $N=10$ particles. }
 
\label{fig1:vapT}
\end{figure}

The results discussed below are obtained for a system of $\Omega=10$ doubly-folded equidistant
levels whose energies are
 \begin{equation}
 \varepsilon_i=\left( i-\frac{1}{2}(\Omega+1) \right)\Delta\varepsilon,~~~~~~~~i=1,.~.~.~.,\Omega 
\end{equation}
and a pairing strength $G=0.4\Delta \varepsilon$. In the following, the total
energy, pairing gap and the temperature are given in units of  $\Delta
\varepsilon$. To take advantage of the $U(1)$ symmetry breaking, the hamiltonian
$\hat h$ is written as a sum of quasi-particle excitations $\hat h = \sum_{k}
E_k \hat \alpha^\dagger_k \hat \alpha_k$, where the $E_k$ denotes the
eigenvalues of the underlying HFB hamiltonian, while the quasi-particle creation
operators write 
\begin{equation}
\hat \alpha^\dagger_k = u_k \hat c^\dagger_{k,+} - v_k c_{k,-},~~ \hat \alpha^\dagger_{\bar k} = u_k \hat c^\dagger_{k,-} + v_k c_{k,+} .
\end{equation} 
Similarly to what is done in nuclear physics,  two levels of
complexity exist in the application of projection techniques. The projection can be
made either before (Variation After Projection [VAP]) or after (Projection After
Variation [PAV]) variation\cite{Rin80}. The latter is much less demanding since
it only requires to solve finite temperature BCS (FT-BCS) equations and make
projection without minimizing Eq. (\ref{eq:varia}). As an illustration, the
temperature dependence of the energy $\langle E \rangle$ and the associated heat
capacity defined through $C_V = {\partial \langle E \rangle} /{\partial T}$
obtained with FT-BCS (dashed line) and FT-PAV (dotted line) are compared to the
exact result (thick line) in figure \ref{fig1:vapT} for $N=10$ particles.
The exact solution is obtained following ref. \cite{Sum07}.
As it is well know, in addition to the systematic overestimation of the energy,
the FT-BCS theory suffers from the sharp superfluid to normal phase transition
as the temperature increases. On opposite, the exact solution display a much
smoother behaviour. It is clearly seen in this figure that, except in the very
small temperature case, the FT-BCS+PAV does even a worse job and does not cure
the threshold effect.

Extrapolating the improvement generally observed at $T=0$ \cite{Hup11} to the
finite temperature case, one can anticipate a much better description if VAP is
performed. In that case, the variational principle (\ref{eq:varia}) should be
minimized by both varying the components $(u_k,v_k)$ and the energy $E_k$
consistently \cite{Ese93}. While in principle possible, such minimization has
never been performed due to the fact that the hamiltonian $\hat h$ and the
operator ${\hat P}_N \exp(-\beta \hat h)  {\hat P}_N $ do not commute and
therefore cannot be diagonalized simultaneously. As a consequence, while a
guideline of practical implementation has been proposed long ago in
\cite{Ese93}, except in the case of the two-level degenerate system, the
predictive power of VAP at finite temperature (called hereafter FT-VAP) has
never been attested. 
      
In the present work, we applied the FT-VAP following the strategy proposed in
ref.  \cite{Ese93}. In practice, the variational principle is minimized by
writing first the energy in terms of the one- and two-body density of the
projected density, both of them written as a non-trivial function of the $u_k$,
$v_k$ and $E_k$ (see Eq. (36) in ref.  \cite{Ese93}).   The minimization is
carried out via a sequential quadratic programming method by using the $v_k$ and
$E_k$ as variational parameters. To compute the free energy without
approximation, at each iteration of the minimization, the entropy is calculated
by
\begin{eqnarray}
S=-k_B \sum_i D^N_i log D^N_i
\end{eqnarray}
where $D^N_i$ are the eigenvalues of the statistical operator $\exp(-\beta \hat
h)/Z$ in the Fock space composed by all the many-body configurations with $N$
particles. Each configuration is characterized by $\eta$ pairs and $I$ unpaired
particles, with 2$\eta+I=N$. Moreover, since states with a different number of
unpaired particles cannot be connected by the operator $\exp(-\beta \hat h)$,
the problem is reduced to the diagonalization of block matrices for each allowed
seniority I. The required computational cost is thus given essentially by two
operations, i.e. the calculation of the matrix elements of the statistical
operators and the diagonalization itself. For the latter, a standard QR
algorithm is used. The calculation of the matrix elements is done by
using the bit representation of the many-body states (see for example ref.
\cite{Caurier}). Each configuration is identified by an integer word whose bits
correspond to the single particle levels and have value 1 or 0 depending on
whether the level is occupied or empty. In such a way all the matrix elements can
be obtained by using very simple logical operations which allow to perform
calculations much faster. 

\begin{figure}[htbp]  
\includegraphics[width=8.cm]{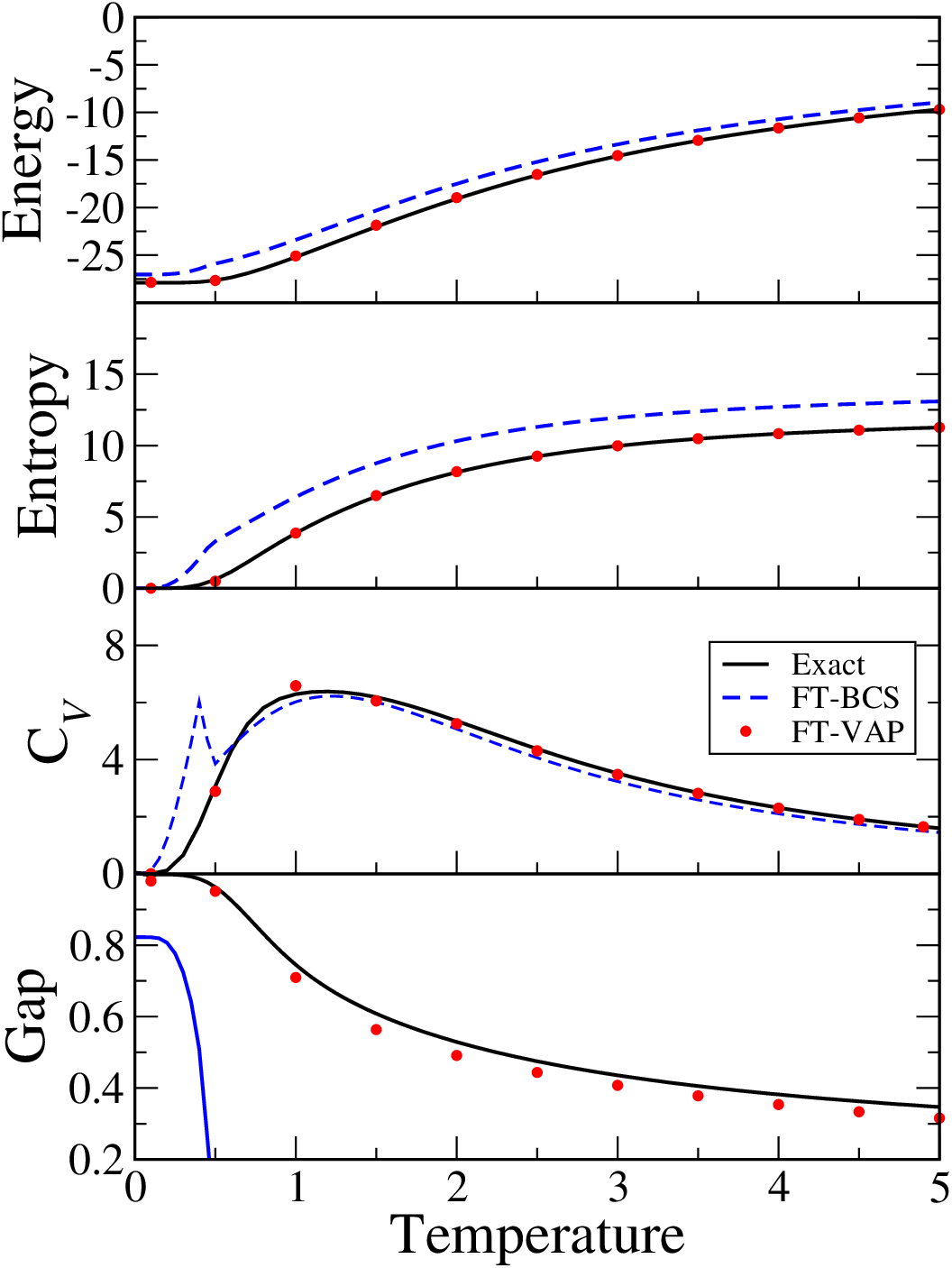}  
\caption{Illustration of the predictive power of the FT-VAP. From top to bottom,
the energy $\langle H \rangle$, the entropy, the heat capacity 
and the average gap obtained with the FT-VAP (filled circles) are compared to
the exact (thick line) and FT-BCS case (dashed line) for a system of $N=10$ particles.}  
\label{fig2:vapT}  
\end{figure}

In figure \ref{fig2:vapT}, the result obtained in FT-VAP is compared to the
exact solution for a system of $N=10$ particles at various temperature. In
FT-VAP, the gap is given by Eq.  (42) of ref. \cite{Ese93} while
in the exact case it is computed through:
\begin{eqnarray}
\Delta=\sqrt{-G (E-E_{0}})
\end{eqnarray}
where $E$ is the total exact energy and $E_0$ is given by  
\begin{eqnarray}
E_{0}=\sum_i \left(\varepsilon_i-\frac{G}{2} n_i\right)n_i
\end{eqnarray} 
containing both the single-particle and the self-energy terms, $n_i$ being the
occupation numbers. In this figure, we see that, except for the small systematic
difference observed for the gap, the FT-VAP approach provides a perfect
description of the thermodynamics of a system with fixed particle number in any
range of temperature. None of the limitations \cite{Ros97,Din08} appearing in
other mean-field based theories are seen. In particular, the entropy, that is an
approximation in FT-VAP,  perfectly matches the exact one. The same quality of
agreement is found also at higher temperature (up to T= 10).

\begin{figure}[htbp]  
  \includegraphics[width=8.cm]{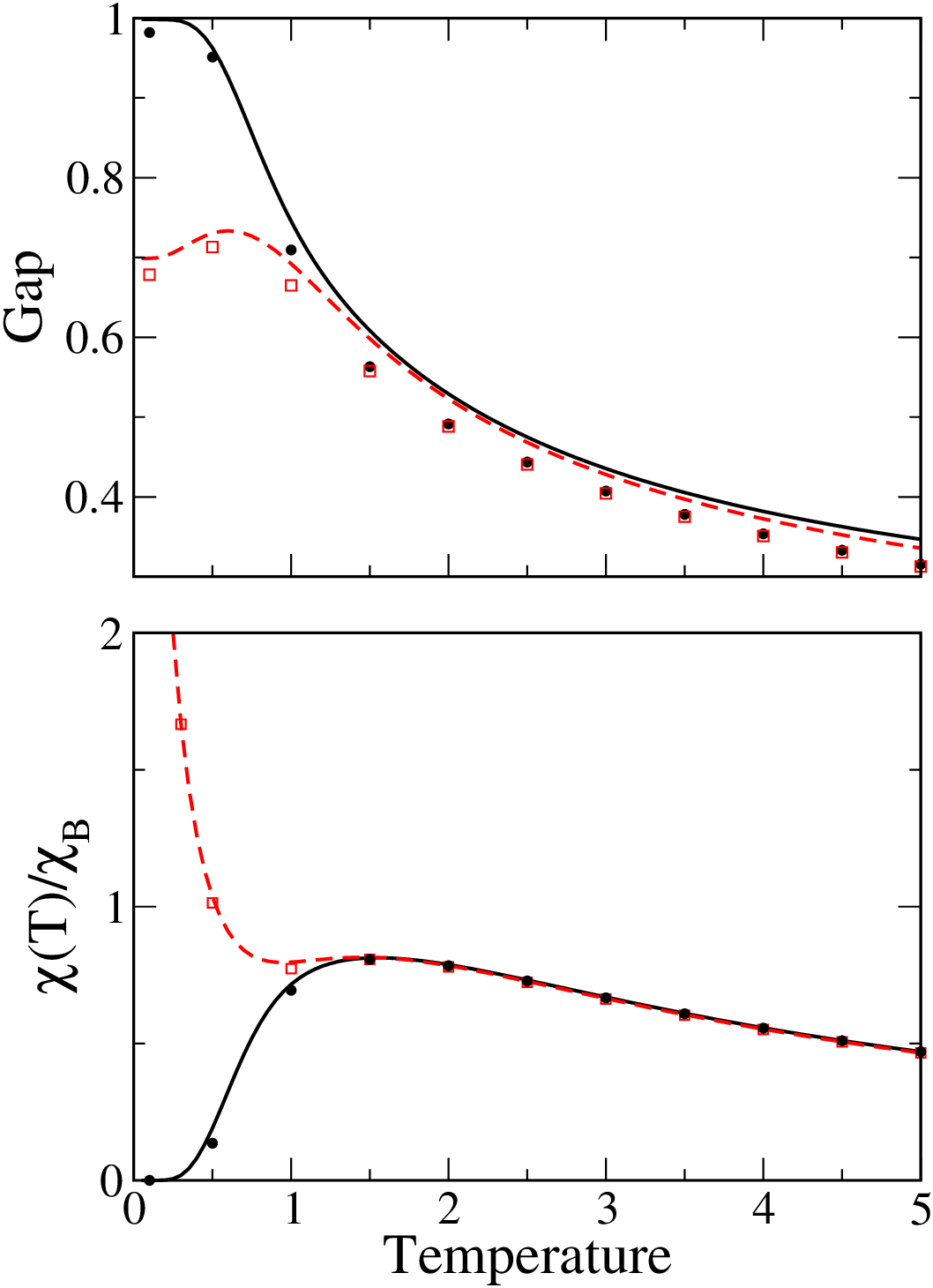}
  \caption{Evolution of the mean gap (upper panel ) and spin susceptibility (lower panel) 
  for $N=10$ (filled circles) and $N=11$ (open squares) particles as a function
of $T$ obtained   with the  FT-VAP. The corresponding exact result for the even and odd systems
are presented respectively with solid and  dashed lines. In this figure, the spin susceptibility is normalized by the
bulk high temperature value $\chi_B = 2 \mu^2_B / \Delta \varepsilon$.}  
  \label{fig3:vapT}  
 \end{figure}

We further investigated the applicability of FT-VAP for odd number of particles.
Taking advantage of the fact that the FT-BCS density mixes up odd and even parities
as soon as a non-zero temperature is applied, we used the same technique as in
the even case. The only difference is that now the projector entering in the
density (Eq. (\ref{eq:dn})) corresponds to an odd number of particles. In top
panel of figure \ref{fig3:vapT}, the pairing gap obtained in FT-VAP for $N=10$
and $N=11$ particles is compared to the exact case. In bottom panel of this
figure, the spin susceptibility $\chi$ defined as \cite{Dil00}  

\begin{eqnarray}
\chi (T)&=& - T \left. \frac{\partial^2 \ln{Z} }{\partial B^2}\right|_{B=0},
\end{eqnarray}     
is shown for the two cases.

In the limit of small magnetic field, the susceptibility identifies with the fluctuation of the magnetization $\hat M = -\mu_B \sum_{\sigma,i} \sigma
\hat c^\dagger_{i,\sigma}\hat c_{i,\sigma} $ \cite{Van06}, i.e. 
\begin{eqnarray}
\chi (T)&=& - \frac{1}{T} \left( \langle \hat M^2 \rangle - \langle \hat M\rangle^2 \right). 
\end{eqnarray}    
In small systems, large differences are observed in the thermodynamics of odd and even systems \cite{Von01}. This is clearly seen 
especially 
at low temperature for the gap. The spin susceptibility 
further underlines the  differences. The FT-VAP perfectly grasps the thermodynamics of odd systems and one cannot distinguish its result 
from the exact solution. 

In the present letter, we applied for the first time the variation after
projection approach to describe the canonical properties of a superconducting
system.  The minimization of the free energy is made with no approximation on
the entropy. The FT-VAP provides a perfect reproduction of the exact result in
the Richardson Hamiltonian case both in the low and high temperature limit and
does not have the limitation of other mean-field based approaches. Due to the
necessity to make use of explicit diagonalization for the entropy, the present
approach is still restricted to rather small number of particles. Nevertheless,
we believe that the result obtained here is sufficiently promising that in the
near future, an effort should be made to promote the FT-VAP and make it more
versatile. It should be mentioned that the present method provides a natural
extension of the FT-BCS or FT-HFB theory presently used to describe nuclei
within the Energy Density Functional framework applied at finite temperature
\cite{edf}.
\section*{Acknowledgement}  
We would like to thank J. Margueron, E. Khan and N. Sandulescu 
for discussion at the early stage of the project. 
  
\end{document}